\documentstyle[preprint,aps]{revtex}
\def\gaas{GaAs/AlGaAs }
\def\H{H_{\|}}
\def\bqn{\begin{equation}}
\def\nqn{\end{equation}}
\begin{document}
\draft
\preprint{}
\title{A New Liquid Phase and Metal-Insulator Transition in Si MOSFETs }
\author{Song~He}
\address{
Bell Laboratories, Lucent Technologies, Murray Hill, NJ 07974 \\
}
\author{X.C.~Xie}
\address{
Department of Physics\\
Oklahoma State University\\
Stillwater, OK 74078 \\
}

\date{August 26, 1997}
\maketitle
\begin{abstract}
We argue that there is a new liquid phase in the two-dimensional electron system in Si MOSFETs at low enough electron densities. 
The recently observed metal-insulator transition results as a crossover from the percolation transition of the liquid phase 
through the disorder landscape in the system below the liquid-gas critical temperature. The consequences of our theory are discussed for variety of 
physical properties relevant to the recent experiments.
\end{abstract}

\narrowtext
The scaling theory of localization\cite{Gang} of non-interacting electrons tells us that disorder is a relevant
perturbation in two-dimension: The system is an insulator at low enough temperatures. The recent
discovery of a possible metal-insulator transition in Si MOSFETs by Kravchenko {\it et al.}\cite{Krav1} apparently
defies this long-held belief. This should not be surprising because the dominant Coulomb interaction 
in Si MOSFETs may invalidate the non-interacting scaling theory. 
In a typical Si MOSFET sample where the transition
is observed, the ratio between the average Coulomb interaction and the Fermi energy is about $20$.
These intriguing experiments\cite{Krav1,Pop} generate
renewed interests in the effects of interaction in disordered systems\cite{Eli}.
One of the striking features of the system is the I-V nonlinearity at average electric fields much weaker than the expected value
determined by the effective temperature of the electrons\cite{Krav2}. 
This raises questions of whether the system can be described by a theory
for homogeneous systems. (See the discussion on electric field
dependence in the text for more detail.)
As we argue in this letter, the mobile positive background in Si MOSFETs allows
macroscopic inhomogeneity to occur more easily at low enough average electron densities. The metal-insulator 
transition, the small field nonlinear I-V, as well as a host of 
other phenomena result from the combined effect of the 
inhomogeneity and the disorder in the system.

In a remote doped or modulation doped GaAs/AlGaAs sample, the electron system is well described by the classic jellium model 
of electron gas. The roughly uniform positive background is fixed by the bulk crystal and it does not participate in the 
dynamics of the electron system. On the other hand, in an inverted Si MOSFET, the positive background charges at the 
metal-oxide interface are mobile. They couple to the dynamics of the electron system, as exhibited by the recent effective 
mass measurements in such systems\cite{Dan1}. This allows for more possibilities for the physical properties 
of the electron system in Si MOSFETs. However, we want to emphasize that the mobile positive background is not absolutely 
essential for the macroscopic inhomogeneity to occur. A slowly-varying disorder potential combined with electron-electron 
correlation, for example, may also favor macroscopic inhomogeneity in the system. 

We now consider what happens at low enough electron densities in a Si MOSFET. For the convenience of our argument, 
we treat the mobile positive background charges as holes. We consider the adiabatic evolution of the system as the oxide 
thickness increases. At zero thickness when the electrons and 
holes are at the same plane, the phase diagram of such an electron-hole system, shown in Fig.\ \ref{Fig1}, is well
established\cite{Moris1,Gordon1}. The high density phase is a metallic electron-hole liquid.
The low density phase is an exciton gas. There is a two-phase coexistence region below the critical temperature $T_c$. 
As the oxide thickness 
increases beyond the Bohr radius $a_B$, 
the excitons in the gas phase dissolve into 
separated electrons
and holes. In a disordered sample, these electrons 
occupy the lowest lying localized states in the system. 
Because these states are strongly localized, the internal degrees of freedom, such as spin and valley indices, do not play
important roles in determining the overall energy of the many-body system in the gas phase. This implies that the energy cost to
polarize these internal degrees of freedom is small compared to the cohesive energy of the liquid phase.
Since there is no excitonic correlation in the electron-hole liquid phase, the length scale set by the Bohr radius does not have a 
special 
meaning to the liquid. As long as the oxide thickness is comparable to the inter-electron distance, the liquid 
phase remains intact except that its cohesive and surface energies 
are reduced. The liquid can be viewed as an entanglement of two normal Fermi liquids of 
electrons and holes\cite{Moris1}. Its internal degrees of freedom couple strongly to the dynamics of the system. 
It is a singlet of the spin and valley degrees of freedom. 
In a disordered system, the liquid phase further lowers its energy by
occupying the valleys of the disorder landscape. It percolates 
through the sample in a fashion determined by the competition among
its cohesive energy, its surface energy, and the disorder potential. This gives rise to a metal-insulator 
transition at the percolation threshold in the classical limit at zero temperature. 
Apparently, quantum tunneling 
between separated liquid regions or a finite temperature destroys this transition. This is illustrated in 
Fig.\ \ref{Fig2}. 
Along the $n$-axis, {\it i.e.}\ at zero temperature and when the coupling constant determining the phase breaking 
mechanism is infinity, there is a second order phase transition at the percolation threshold $n=n_c$. 
Away from the $n$-axis, the transition is 
destroyed by a finite temperature or quantum tunneling. 
However, in the region of temperature and coupling constant close to 
the critical point, the physics of the system is influenced by 
the very existence of the critical point, similar to the situation in 
a quantum phase transition\cite{Subir1,Senthil}. Therefore it is the crossover behavior
near the critical point that is observed in the experiments.

We now discuss the related experiments in Si MOSFETs and show how 
they are consistent with the consequences of our proposed theoretical 
framework.

{\em Temperature dependence of the resistivity}---One of the most striking 
features of the experiments\cite{Krav1,Pop} is that the temperature dependence
of the resistivity changes from insulator-like to metal-like as the 
average electron density increases. This is readily understood
in our theory. At temperatures below the critical temperature $T_c$ and 
at a low enough average density, the system is in the two-phase
regime. When the average density is below the threshold density of the percolation, the conduction 
in the system is dominated by the phonon-assisted hopping through
the localized gas phase. At low enough temperatures, this gives the 
famous Coulomb-gap behavior\cite{Cgap}
\bqn
\rho\sim e^{\sqrt{E_{0}/T}},
\nqn
where $E_{0}$ is an energy scale determined by the localization 
length and the Coulomb interaction.
On the other hand, when the average density is above the threshold density, 
the metallic liquid phase percolates through the whole
sample. The sample is able to conduct electric current. Using a simple two-fluid model for the conduction, we have
\bqn
\sigma=f_l\sigma_l+f_g\sigma_g \mbox{~~~and~~~} f_l+f_g=1,
\nqn
where $f_l$ and $f_g$ are respectively the fractions of the electrons in the liquid and the gas phases.
$\sigma$, $\sigma_l$, and $\sigma_g$ are respectively the total conductivity, 
the conductivity of the liquid and the conductivity of the localized gas. At temperatures much lower than the cohesive energy
$\Delta_c$ of the liquid, we estimate $f_g$ to be
\bqn
f_g\simeq Ae^{-\frac{\Delta_c}{T}}.
\nqn
Combining the above equations, we obtain the low temperature 
behavior of the resistivity on the metallic side
\bqn
\label{RT}
\rho=\tilde{\rho}_0+\tilde{\rho}_1e^{-\frac{\Delta_c}{T}},
\nqn
where $\tilde{\rho}_0=1/\sigma_l$ and $\tilde{\rho}_1=A\tilde{\rho}_0(1-\sigma_g/\sigma_l)$. 
This temperature dependence of $\rho$ on the metallic side was first suggested by Pudalov\cite{Pudalov} 
based on completely different interpretation of the experiments. It was
implicit in the experimental data of Kravchenko {\it et al.}\cite{Krav1}. It was explicitly 
observed by Hanein {\it et al.}\cite{Dan2} in recent experiments 
on \gaas hole samples with a conducting backgate. 
For $T>\Delta_c$, the liquid phase ceases to exist. The resistivity of the system depends weakly on temperature.

{\em Scaling and duality}---In a realistic sample, the coupling 
constant of the phase breaking mechanism is fixed by the property of the sample.
We need to consider what happens in a constant-$\alpha$ plane in Fig.\ \ref{Fig2}. 
There is a diverging length scale $\xi$ and a vanishing energy scale 
$\varepsilon\sim e^2/\xi$ as the system approaches the critical point $n_c$. Using the usual arguments\cite{Girvin1}, 
in the region close to the critical point, we write the finite temperature resistivity of the system in the scaling 
form
\bqn
\rho(T,n)=\tilde{\rho}(\delta/T^{1/\nu}),
\nqn
where $\delta=(n-n_c)$ and $\nu$ is the correlation length exponent. 
The value of $\nu$ predicted by the 
percolation theory is $\nu=4/3$. This is within the range of the measured values of $1.2\sim 1.6$.

The duality observed in the experiments\cite{Krav1.5,Pop} can be readily 
understood by the following argument.
Because there is no critical point on the constant-$\alpha$ plane at a 
finite temperature, the function $\tilde{\rho}(x)$ 
is an analytic function of its variable. Taylor expanding it at $x=0$, we have
\bqn
\rho(T,n_c+\delta)=\tilde{\rho}^{(0)}+\tilde{\rho}^{(1)}\frac{\delta}{T^{1/\nu}}+\frac{1}{2}\tilde{\rho}^{(2)}(\frac{\delta}{T^{1/\nu}})^2+\cdot\cdot\cdot,
\nqn
where $\tilde{\rho}^{(0)}=\tilde{\rho}(0)$, $\tilde{\rho}^{(1)}=\frac{d\tilde{\rho}(x)}{dx}\mid_{x=0}$, 
and $\tilde{\rho}^{(2)}=\frac{d^2\tilde{\rho}(x)}{dx^2}|_{x=0}$. Immediately we have
\bqn
\rho(T,n_c+\delta)\rho(T,n_c-\delta)=(\tilde{\rho}^{(0)})^2+\mbox{O}(\delta^2).
\nqn
Obviously, this operationally defined duality relation is a generic feature of the critical region. 
It should be interesting to 
test this by looking at the temperature dependence of the coefficient of the $\delta^2$ term.

{\em Electric-field dependence of the resistivity}---There are two
important features associated with the electric-field experiments\cite{Krav2}.
First, the nonlinear I-V occurs at very small current (or electric field).
Second, the nonlinear resistivity exhibits scaling as a function
of electric field in both the metallic and insulating sides. 
There are two known mechanisms through which nonlinearity can occur in a homogeneous system\cite{Girvin1}. 
The first one requires that the 
energy scale determined by the electric field to be the dominant energy scale in the system.
In the critical region, this energy should be larger than
the temperature. However, in a typical nonlinear I-V experiment in Si 
MOSFETs, the nonlinearity occurs at electric fields as low as
$0.25$ mV/cm\cite{Krav2}. Using a phase breaking length of $10000$\AA, we estimate the 
nonlinear electric field energy scale to be about $0.3$mK,
much smaller than the temperature $220$mK. No nonlinearity should result from this mechanism.
The second mechanism is due to Joule heating. Joule heating raises the effective temperature $T_{e}$ of the electrons
according to $T_e\sim E^{1/2}$\cite{Chow}.
The experiments in Ref.\cite{Krav2} indicate that $T_e$ is about $2$K at an electric field of $50$mV/cm. Using an electric field
of $0.25$mV/cm, we estimate $T_{e}$ to be $150$mK. This is lower than the lattice temperature $220$mK. 
Therefore Joule heating is not significant at such an electric field and no nonlinearity should result. 
The above arguments leads us to believe that the nonlinear I-V is caused by the intrinsic inhomogeneity in the system.

The scaling behavior appeared at large electric fields can be readily
understood following the arguments in Ref.\cite{Girvin1}. Close
to the critical point, the vanishing energy scale $\epsilon \sim e^{2}/\xi$
is cut off by the electric field through $\epsilon \sim eE \xi$. Thus,
the zero temperature resistivity can be written in the scaling form
\bqn
\rho(E,n)= \tilde{\rho}(\delta/E^{1/\nu(z+1)}),
\nqn
where $z=1$ because the vanishing energy scale at the critical point is given by $\epsilon \sim e^{2}/\xi$. Using the 
classic value of $\nu=4/3$, we find $\nu(z+1)=8/3\simeq 2.66$, which is very close to the experimental value of $2.70$\cite{Krav2}.

{\em Effects of an in-plane magnetic field}---When the Zeeman energy of the magnetic field is larger than the Fermi energy of the 
gas phase, it fully polarizes the electrons in the gas phase. The portion of electrons in the gas phase at low temperatures 
can be estimated to be
\bqn
f_g\sim e^{\frac{g\mu_BS_zH_{\|}-\Delta_c}{T}}
\nqn
when $g\mu_BS_zH_{\|}<\Delta_c$. Using the two-fluid model above, we find the in-plane magnetic field dependence of the resistivity on the metallic side
at low temperatures to be 
\bqn
\label{H}
\rho\sim\rho_0+\rho_1e^{g\mu_BS_zH_{\|}/T}
\nqn
When the Zeeman energy goes beyond $\Delta_c$, the liquid phase ceases to exist because the critical temperature 
$T_c$ is zero. The system is a fully polarized gas and it is not affected by further increases of $\H$. 
Fig.\ \ref{Fig3} shows this generalized situation of Eqn.(\ref{H}) for a given temperature $T$.
The resistivity increases exponentially with $\H$ from its zero-field value $\rho_{0}(T)$ to the saturated value $\rho_{m}(T)$ at 
the critical field $H_{\|,c}$ given by $g\mu_BS_zH_{\|,c}=\Delta_c$, {\it i.e.} 
\bqn
\label{HH}
\ln\rho(T,\H)=\left\{ \begin{array}{ll}
\ln\rho_0(T)+\frac{\H}{H_{\|,c}}(\ln\rho_m(T)-\ln\rho_0(T)) & \mbox{if $\H<H_{\|,c}$}\\
\ln\rho_m(T) & \mbox{if $\H>H_{\|,c}$}
\end{array}\right.
\nqn
When $T\ll g\mu_BS_zH_{\|,c}$, $\rho(T,\H)$ should be exponential in $\H/T$ for intermediate values of $\H$. Using this we have
\bqn
\label{RM}
\ln\rho_{m}(T)\sim\ln\tilde{\rho}_m+\frac{g\mu_BS_zH_{\|,c}}{T}.
\nqn
The preliminary experimental data\cite{Krav3} is consistent with this prediction. 
It should be interesting to test it in more details.

Next we look at how the low temperature behaviors of the resistivity on the metallic side evolve for different in-plane magnetic
fields. Using Eqn.(\ref{HH}) and Eqn.(\ref{RM}), we have
\bqn
\ln\rho(T,\H)=(1-\frac{\H}{H_{\|,c}})\ln\rho_0(T)+\frac{g\mu_BS_z\H}{T}+\frac{\H}{H_{\|,c}}\ln\tilde{\rho}_m.
\nqn
This equation predicts different behaviors of $\rho(T,\H)$, shown in the inset of Fig.\ \ref{Fig3}, for $\H<H_{\|,0}$ and $\H>H_{\|,0}$, where
$H_{\|,0}=\tilde{\rho}_1H_{\|,c}/(\tilde{\rho}_0+2\tilde{\rho}_1)$. Using the experimental data of 
$\tilde{\rho}_1/\tilde{\rho}_0\simeq 10$ and $H_{\|,c}\simeq 20$ kOe
in the experiments of Simonian {\it et al.}\cite{Krav3}, 
we predict $H_{\|,0}\simeq 10$ kOe. This is consistent with the data
shown in Figure 4 in Ref.\cite{Krav3}.
For $\H\ll H_{\|,0}$, the temperature $T_m$ at which $\rho(T,\H)$ reaches a minimum is given by
\bqn
T_m\sim\frac{\Delta_c}{\ln(\frac{\Delta_c}{g\mu_BS_z\H})}.
\nqn
This is consistent with the preliminary experimental data\cite{Krav3}. 
It should be interesting to test this prediction quantitatively.

{\em Further experiments}---(1) One of the most interesting experiments is to probe directly the inhomogeneity in 
the electron density. This can be achieved through scanning techniques using a single-electron-transistor as the probe
for the electric field generated by the electrons in the two-dimensional electron system. 
According to our theory, we expect the system to phase separate into regions of different densities at low enough average 
electron densities. The regions of the sample occupied by the liquid phase should be easily observable in one of the implementations
of such scanning techniques\cite{Amir}. (2) In our theory, the spin and valley indices play similar roles in determining the physics
of the system. We expect a strain field, which splits the valley degeneracy of the two-dimensional electron
system in Si MOSFETs, to have effects similar to those of an in-plane magnetic field. This should distinguish our theory
from those in which the spin degrees of freedom play a non-trivial role in producing the observed experiments. (3) Light
scattering experiments in such systems can also provide useful information. Because of the occurrence of macroscopic inhomogeneity in the 
system below the critical temperature $T_c$, we should see an increase in the wave-vector-breaking components of the light scattering as the
temperature gets below $T_c$.

In summary, we have argued that there is a new liquid phase in the two-dimensional electron system in Si MOSFETs at low enough
average electron densities. The percolation transition of this liquid through the disorder landscape in the classical limit 
in a disordered system causes the metal-insulator crossover behaviors observed in recent experiments. We have made quantitative 
predictions and proposed new experiments to further investigate the physical properties of the system.

\noindent Acknowledgments: We thank S.~V.~Kravchenko, D.~Simonian, D.~C.~Tsui, Senthil Todadri, Steve Simmon, and X.~R.~Wang for stimulating
discussion. We also thank the authors of Ref.\cite{Girvin1} for doing the community a big favor by writing such a beautiful
article. X.~C.~Xie is supported by the NSF-EPSCOR program.

\begin{figure}
\caption{Phase diagram of an electron-hole system.$~~~~~~~~~~~~~~~~~~~~~~~~~~~~~~~~~~~~~~~~~~~~~~~~~~~~~$}
\label{Fig1}
\end{figure}

\begin{figure}
\caption{Effects of a finite temperature and quantum tunneling on the percolation transition.$~~~~~~~~~~~~~~~$}
\label{Fig2}
\end{figure}

\begin{figure}
\caption{In-plane magnetic field dependence of the resistivity. The Inset shows the temperature dependence of $\rho$ for
various in-plane magnetic fields: (a) $H_{\|}\ll\tilde{H}_{\|,0}$, (b) $H_{\|}<\tilde{H}_{\|,0}$, 
(c) $H_{\|}>\tilde{H}_{\|,0}$. The temperature is measured in unit of $g\mu_BS_zH_{\|,c}$.}
\label{Fig3}
\end{figure}

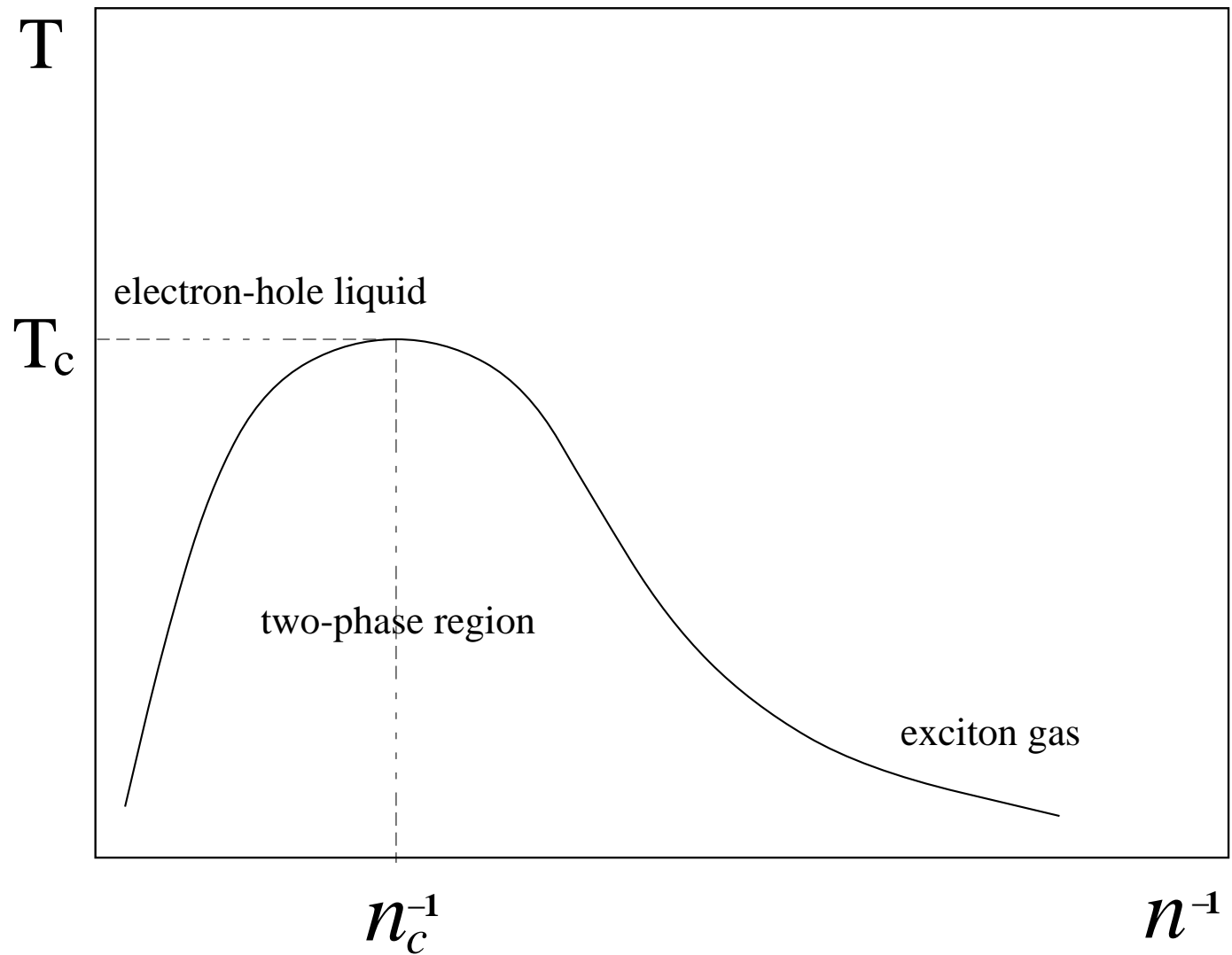
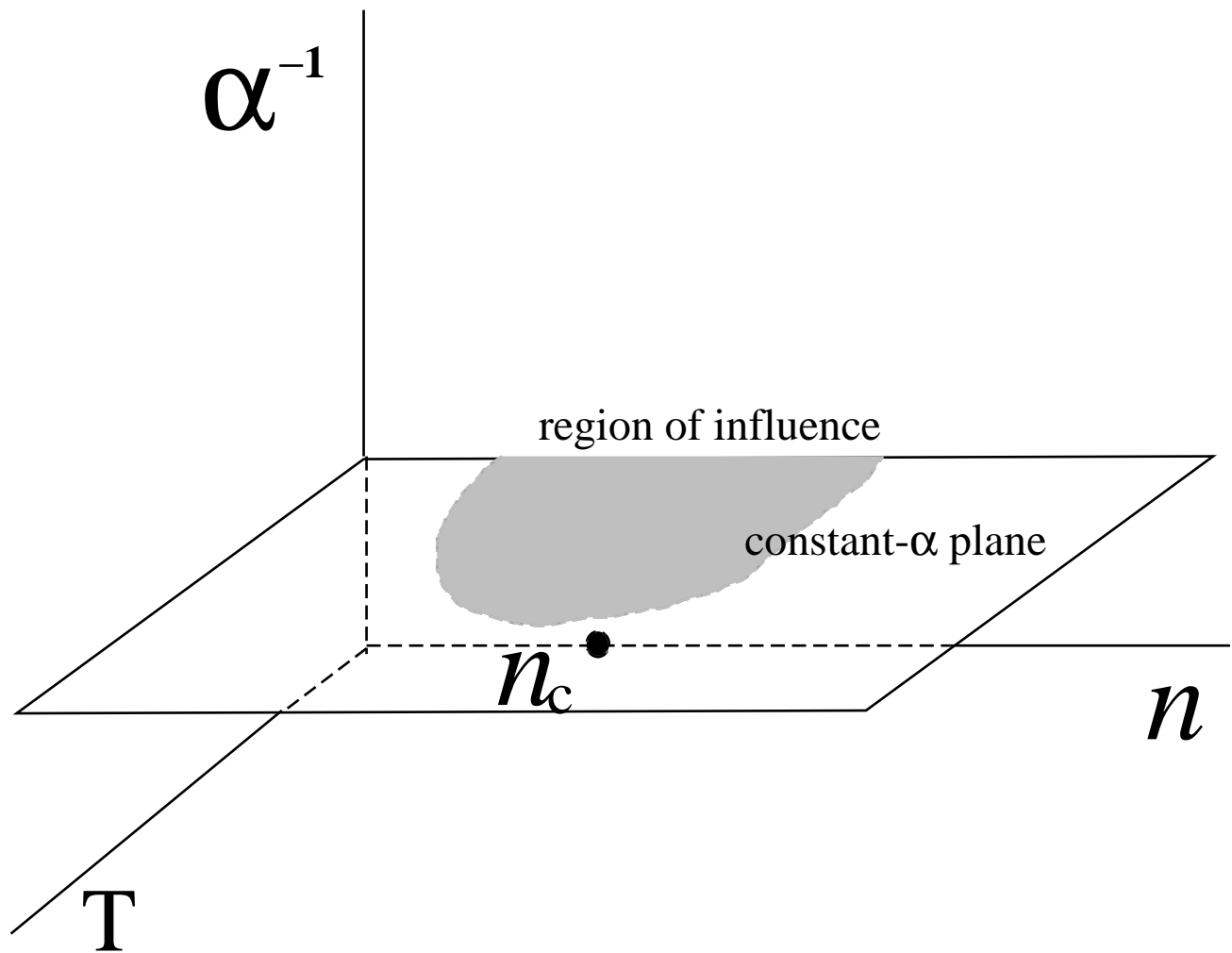
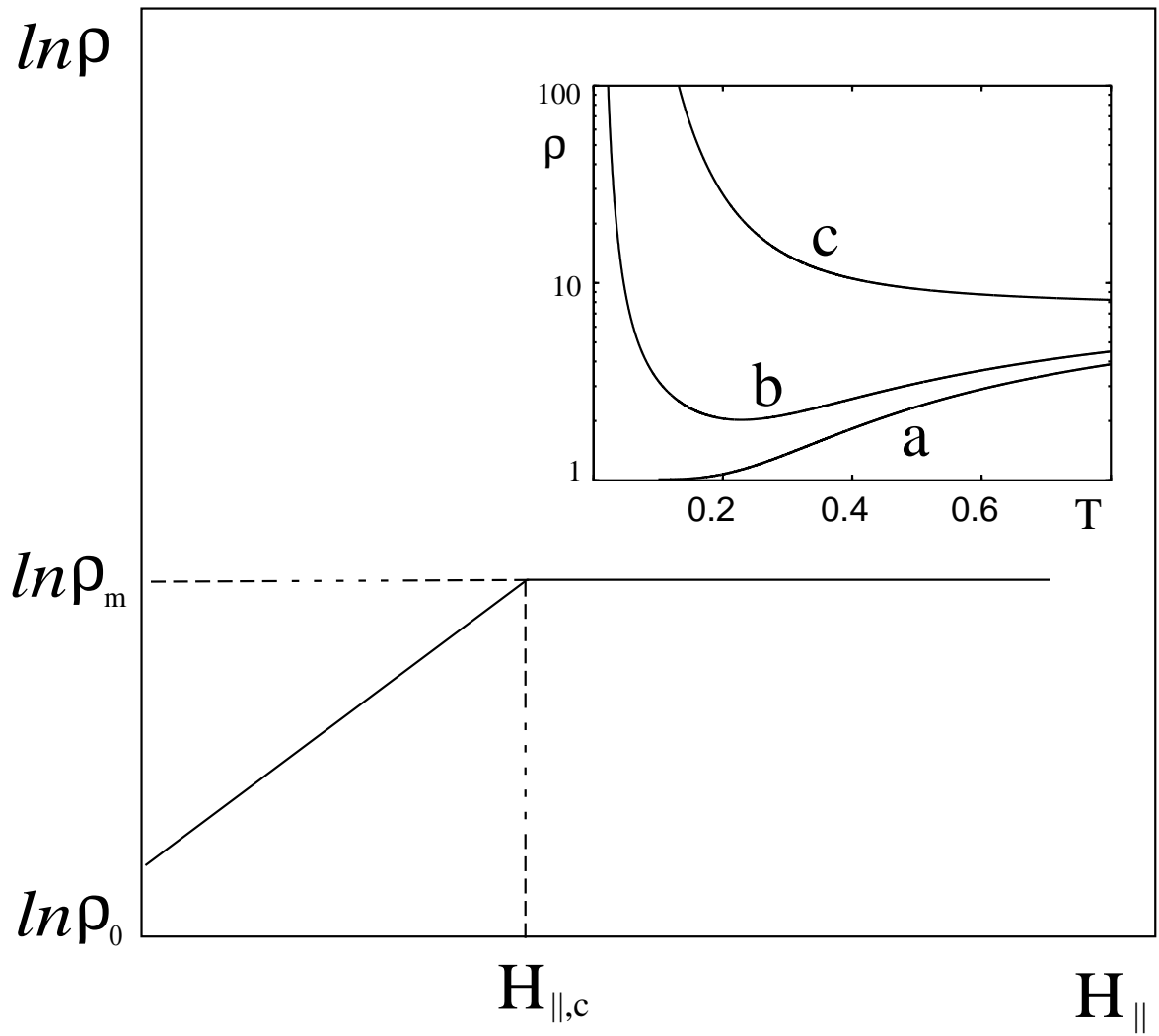
\end{document}